\documentclass[preprint,12pt]{elsarticle}

\usepackage{amssymb}
\journal{Planetary and Space Sciences}

\begin{document}

\begin{frontmatter}



  \title{Asteroid families interacting with secular resonances}


\author{V. Carruba$^{1}$, D. Vokrouhlick\'{y}$^{2}$, B. Novakovi\'{c}$^{3}$}
\ead{vcarruba@gmail.com,vokrouhl@cesnet.cz,bojan@matf.bg.ac.rs}
\address{$^{1}$S\~{a}o Paulo State University (UNESP), School of Natural Sciences and Engineering, Guaratinguet\'{a}, SP, 12516-410, Brazil}
\address{$^{2}$Institute of Astronomy, Charles University, V Hole\v{s}ovi\v{c}k\'{a}ch 2, Prague 8, CZ-18000, Czech Republic}
\address{$^{3}$University of Belgrade, Faculty of Mathematics, Department of Astronomy, 
Studentski trg 16, 11000 Belgrade, Serbia}

\begin{abstract}
  Asteroid families are formed as the result of collisions. Large fragments 
  are ejected with speeds of the order of the escape velocity from the
  parent body. After a family has been formed, the
  fragments' orbits evolve in the space of proper elements because of
  gravitational and  non-gravitational perturbations, such as the Yarkovsky
  effect.  Disentangling  the contribution to the current orbital position of
  family members caused by the initial ejection velocity field and the
  subsequent orbital evolution is usually a difficult task. Among the more
  than 100 asteroid families currently known, some interact with linear and
  non-linear secular resonances.  Linear secular resonances occur when there
  is a commensurability between the precession frequency of the longitude of
  the pericenter ($g$) or of the longitude of node ($s$) of an asteroid and
  a planet, or a massive asteroid.  The linear secular resonance most
  effective in increasing an asteroid eccentricity is the ${\nu}_6$, 
  that corresponds to a commensurability between the precession
  frequency $g$ of an asteroid and Saturn's $g_6$. Non-linear secular
  resonances involve commensurabilities of higher order, and can often
  be expressed as combinations of linear secular resonances. This is the case,
  for instance, of the $z_k=k(g-g_6)+(s-s_6)$ resonances. Asteroid families
  that are crossed by, or even have a large portion of their members, in
  secular resonances are of particular interest in dynamical astronomy. First,
  they often provide a clear evidence of asteroid orbit evolution due to the
  Yarkovsky effect. Second, conserved quantities of secular dynamics can be
  used to set valuable constraints on the magnitude of the original ejection
  velocity field. For the case of the ${\nu}_6$ secular resonance, objects in
  anti-aligned (paradoxal) librating states can be prevented to achieve high
  values of eccentricity and remain long-term stable (the case for members of
  the Tina family).  Finally, by changing the value of inclination of family
  members, and, indirectly, of the $v_W$ component of the observed ejection
  velocity field, nodal secular resonances with massive asteroids or dwarf
  planets, such as the $s-s_C$ secular resonance with Ceres, can cause $v_W$
  to become more and more leptokurtic (i.e., more peaked and with larger
  tails than that of a Gaussian distribution). By simulating fictitious
  asteroid families and by requiring that the current value of the Pearson
  kurtosis of $v_W$, $\gamma_2(v_W)$, be attained, independent constraints on
  the value of families ages can be obtained for families affected by these
  kinds of resonances. 
\end{abstract}


\end{frontmatter}


\section{Introduction}
\label{sec: intro}

Among the more than 100 currently known asteroid families some
are characterized by their interaction with linear or non-linear secular
resonances. Asteroid families that are crossed, or even entirely
immersed, in these resonances are of particular interest in dynamical
astronomy. This is because they often provide evidence of long-term
orbital evolution due to non-gravitational (Yarkovsky) forces and/or
can be used to set independent constraints on the magnitude of the
original ejection velocity field.

One issue that has to be first solved when dealing with asteroid families
interacting with secular resonances is the orbital location of these
resonances. \citet{Hirayama_1923} derived secular frequencies from
Laplace-Lagrange linear theory, and noticed that the ${\nu}_6$ secular
resonance (though he did not use this name) was adjacent to the Flora family.
His work was the first, that we know of, to notice the possible interaction
of secular resonances with asteroid families. Values of secular
frequencies for asteroids and planets were also obtained by
\citet{Brouwer_1950} using a linear theory of secular perturbations,
substantially improved by the introduction of the second order terms in
perturbing mass accounting for the 5:2 and 2:1 near commensurabilities
of Jupiter and Saturn.  \citet{Williams_1969} used a non-linear
semi-analytic theory that used the Gaussian method to treat short
periodic perturbations to analyze the orbital evolution
of asteroids over large periods of time.  Two years later, the same
author identified the linear secular resonances in the main belt as
an effective mechanism for the depletion of asteroids.
Our understanding of the dynamics in linear secular resonances greatly
improved later on thanks to analytic models of the ${\nu}_6$ and
other resonances, \citep{Yoshikawa_1987, Knezevic_1991, Morbidelli_1991, 
Morbidelli_1993}, when, using Hamiltonian formalisms, equilibrium points
and the phase space of the resonances were first investigated. 

Knowledge on the orbital position of the asteroid families themselves
advanced significantly in the years 1990s, thanks to a seminal series of 
papers by Milani and Kne\v{z}evi\'{c}, who studied how to analytically
\citep{Milani_1990} and numerically \citep{Knezevic_2000, Knezevic_2003}
determine proper elements, conserved quantities of the motion over timescales of
Myr. These authors also first investigated non-linear secular
resonances \citep{Milani_1992,Milani_1994} and studied
how they affected values of proper
elements, when purely gravitational forces were considered. A modern theory of
proper elements for high-inclination orbits in the main belt, and
the Hungaria region, was initiated by works of Lemaitre and Morbidelli
\citep{Lemaitre_1994a, Lemaitre_1994b}.

It was, however, in the years 2000s, that the dynamical importance
of the Yarkovsky-driven dynamical evolution of asteroids into secular
resonance was first revealed. The Yarkovsky effect is a thermal radiation
force that causes objects to undergo semi-major axis drift as a
function of their size, spin, orbit, and material properties 
\citep{Vokrouhlicky_2015}. The seminal study of \citet{Bottke_2001}
on the Yarkovsky driven mobility of asteroid showed that the shape
of the Koronis family in the $(a,e)$ proper domain can only be explained
if asteroids migrating toward higher semi-major axis interacted
with the pericenter resonance $g+2g_5-3g_6$ and, as a consequence, had
their eccentricity increased because of the passage through the resonance.
Several other papers investigated the role of the interplay of the
Yarkovsky effect with secular resonances, among these one can
quote the work on the Eos family and the $z_1$ secular resonance
(\citet{Vokrouhlicky_2002,Vokrouhlicky_2006a}, interestingly enough
\citet{Brouwer_1951} argued that the Eos family had to be young, because of the
clustering of the secular angles of the then known family members.
\citet{Vokrouhlicky_2006a} showed that this was a consequence of the
Yarkovsky-induced evolution of asteroids into the $z_1$ resonance),
on the Sylvia family and the $z_1$ resonance \citep{Vokrouhlicky_2010},
and of V-type objects and the $z_2$ secular resonance \citep{Carruba_2005}.
\citet{Carruba_2014b} studied the interaction of V-type photometric
candidates with the $g+g_5-2g_6$ secular resonance, a $g$-type resonance,
in particular for the region of the Astraea family.  Recently,
\citet{Milani_2017} obtained resonant proper element adapted to this
resonance and an age estimate for this family, among others.

Of particular interest for the topic of this work was the identification
of the resonant nature of the Agnia \citep{Vokrouhlicky_2006b} and
Tina \citep{Carruba_2011} families. These were the first two families to
have a majority or all members in librating states of the $z_1$ (Agnia)
and ${\nu}_6$ secular resonances.  Conserved quantities of secular dynamics
can be used to set independent constraints on the magnitude of the original
ejection velocity field of these peculiar families.
For the case of the  ${\nu}_6$ secular resonance, objects in anti-aligned
librating states \citep[][or in paradoxal libration,
according to \citet{Ferraz_1985}]{Morbidelli_1991}, 
can be prevented to achieve high values of eccentricity, and remain 
long-term stable, as is the case for members of the Tina family.

Other developments in the 2000s involved the use of frequency
domains such as the $(n,g,g+s)$ for family determination purposes
\citep{Carruba_2007, Carruba_2009a}, and for visualization of secular 
resonances in frequency domains, as also previously done by other
authors \citep{Milani_1994}. The complicated three-dimensional
structure of secular resonances in the $(a,e,\sin i)$ appears as
lines in appropriate frequency domains, making it easier to visualize
and inspect asteroid families affected by these resonances.
The late 2000s also showed that secular resonances with
terrestrial planets are important in the Hungaria
\citep{Warner_2009, Milani_2010} and inner main belt region, for the Vesta
\citep{Carruba_2005} and Phocaea \citep{Zitnik_2015} regions.

Among the most recent highlights was the discovery of the dynamical
importance of secular resonances with massive bodies other than planets,
mostly, but not only, with (1) Ceres \citep{Novakovic_2015, Tsirvoulis_2016}.
The orbital distribution in the $(a,\sin i)$ domain of members of 
the Hoffmeister and Astrid families was for the first time explained
by the effect of nodal resonances $s-s_C$ with Ceres.
Finally, by modifying the value of inclination of family members, and,
indirectly, of the $v_W$ component of the observed ejection velocity field,
nodal secular resonances with massive asteroids or dwarf planets, such as
the $s-s_C$ secular resonance with Ceres, can cause $v_W$ to become more and
more leptokurtic, i.e., more peaked and with larger tails than that of a
Gaussian distribution. By simulating fictitious asteroid families and by
requiring that the current value of the Pearson kurtosis of $v_W$,
$\gamma_2(v_W)$, be attained, independent constraints on the value of
families ages can be obtained for families affected by these kinds of
resonances \citep{Carruba_2016c, Carruba_2016d}.

In this work we briefly review some of the most recent results
on asteroid families affected by secular resonances, with a focus
on asteroid families with a majority of their members inside
a secular resonance and families affected by secular resonances with
Ceres or massive asteroids.

\section{Secular Dynamics}
\label{sec: sec_dyn}

Secular resonances are commensurabilities that involve the frequencies of
the asteroid longitude of perihelion $g$, node $s$ and the fundamental 
frequencies of planetary theories $g_i = \langle\dot{\varpi}\rangle$ and 
$s_i = \langle\dot{\Omega}\rangle$ , where $i$ is a suffix that indicates 
the planets (5 for Jupiter, 6, for Saturn, etc.), and the symbols $\langle
\rangle$ indicate long-term mean value of the corresponding frequency.
Interested readers can look \citet{Knezevic_2000, Knezevic_2003}
for more details on the Fourier transform methods normally used
to numerically obtaining values of planetary and asteroid frequencies.
In order to be in a secular resonance, the proper frequencies have to
satisfy the relationship:

\begin{equation}
  p\cdot g+q \cdot s +\sum_{i}(p_i\cdot g_i +q_i \cdot s_i)=0,
  \label{eq: sec_res}
\end{equation}

\noindent where the integers $p, q, p_i, q_i$ have to fulfill the D'Alembert
rules of permissible arguments: the sum of the coefficients must be
zero and the sum of coefficients of nodal longitudes frequencies
must be even. The combinations (Eq.~\ref{eq: sec_res}) that
only involve the frequency of the asteroid perihelion are often referred to as
“pericenter resonances”, while those with only frequency of the
asteroid node are named as “node resonances”.

\begin{figure}
\centering
\centering \includegraphics [width=0.45\textwidth]{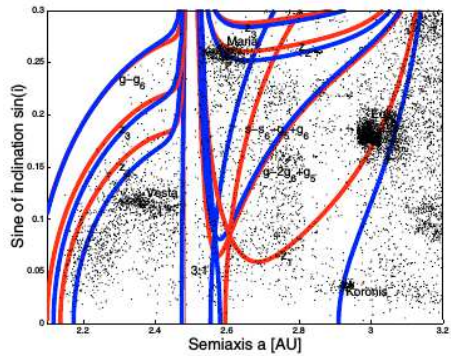}

\caption{Location of some of the main secular resonances in the asteroid belt.
The secular resonances obtained for the fixed value 0.10 of the proper
eccentricity are superimposed to the proper elements $a$ and $\sin i$ of
the asteroids with proper e within the range of $\pm0.025$ of the central value.
The names indicate the main asteroid families in the region. The
pairs of red and blue lines display the locations of the edges of the
resonances. See Figs. 7, 8 and 9 of
\citet{Milani_1994} for more details.}
\label{Fig: sec_res}
\end{figure}

The most important linear secular resonances in asteroid dynamics occur
when ${\nu}_6 = g-g_6 = 0$ arcsec/yr, ${\nu}_5 = g-g_5 = 0$ arcsec/yr, and
${\nu}_{16}= s-s_6 = 0$ arcsec/yr.  Pericenter resonances affect an asteroid
eccentricity, while node resonances influence an asteroid inclination.
Non-linear secular resonances involve higher order commensurabilities, as for
the $z_k = k(g-g_6) +s-s_6$ resonances sequence, that was first investigated by
\citet{Milani_1992,Milani_1994} and later confirmed by
\citet{Vokrouhlicky_2006a,Vokrouhlicky_2006b, Carruba_2009,Carruba_2015},
among others.   While mean-motion resonances have a characteristic
V shape in the $(a,e)$ plane, secular resonances
usually have complicated three-dimensional structures in
the proper $(a, e, \sin i)$ space and cannot easily be drawn
in a plane. Figure~\ref{Fig: sec_res} shows the locations of the ${\nu}_6$
and $z_k$ resonances, of the pericenter resonance $g-2g_6 + g_5$,
and of the node resonance $s-s_6-g_5 + g_6$ determined for proper
$e = 0.10$ (the two
curves refer to resonances computed for values of $e$ to within
$\pm0.025$ of the central value) with method of \citet{Milani_1994} using the
values of planetary frequencies listed in Table~1 of \citet{Carruba_2007}.
The location and shape of the secular resonances change significantly for
different values of the proper $e$, making it difficult to visualize
how asteroid families members may interact with such resonances.

The reason for this is that in the $(a, e, \sin i)$ space the position of a
resonance depends on all three elements. To overcome this
problem, one can project asteroids in different domains, like
that of the proper frequencies $(n, g, s)$, where the mean-motion and secular
resonances are separable\footnote{According to nonlinear perturbation
theories, the proper frequencies are functions of the proper elements
and the transformation from proper elements to proper frequencies is
one-to-one if the Kolgomorov's non-degeneracy condition is
satisfied \citep{Ferraz_2007}.}. In this space, the location of
mean-motion resonances mostly depends on the mean-motion proper frequency
$n$ (see \citet{Knezevic_2000} for a discussion on how to compute this
quantity), while the position of secular resonances only
depends on the proper $(g,s)$ frequencies and appear as lines in a plane. 

\begin{figure}
\centering
\centering \includegraphics [width=0.45\textwidth]{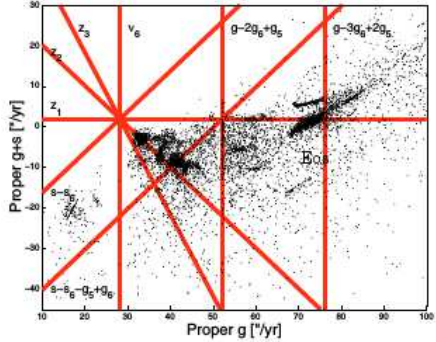}

\caption{A projection of asteroids in the $(g,g+s)$ domain.  The horizontal
  line displays the location of the $z_1 = g-g_6+s-s_6$ secular resonance.
  From Figure 6 of \citet{Carruba_2007}.}
\label{Fig: ggs_plot}
\end{figure}

Different planes can be used to better represent asteroids interacting
with different type of secular resonances.  $g$-type secular resonances
could be better displayed in an $(n,g)$ or $(a,g)$ plane, $s$-type
resonances would be better displayed in $(n,s)$ or $(a,s)$ planes,
while $(g+s)$-type of resonances could be better shown in a plane
in which one of the axis displays $g+s$.  Fig.~\ref{Fig: ggs_plot}
displays a projection of main belt asteroids in the $(g,g+s)$ domain
from \citet{Carruba_2007}.  Other type of representations for
different type of non-linear secular resonances, such as $2g+s, 3g+s,g-s$
and other, were introduced and explored in \citet{Carruba_2009a}.

Asteroids interacting with secular resonances can also be better identified
in frequency domains.  The libration region of each secular resonance has
a typical width around the respective combination of planetary frequencies.
For instance, it was shown that asteroids in librating states of the
${\nu}_6$ secular resonance could be found to within $\pm 1.2$~arcsec/year
from $g = g_6$ for asteroids in the region of the Tina and Euphrosyne
families \citep{Carruba_2011, Machuca_2012}, while the limit for asteroids in
librating states of the $z_1$ resonance was $\pm0.3$ arcsec/year for
$g+s=g_6+s_6$ for asteroids in the Padua family region \citep{Carruba_2009}.
By selecting a cutoff value in frequencies it is therefore possible to
select objects that are likely to be in librating states of a given resonance.
Such objects, or likely resonators, could then be investigated to check
if their resonant argument is actually librating around an equilibrium
point, so as to determine the effective size of the population of objects
in actual resonant configuration.  The likely resonator criteria is, therefore,
an useful tool in pre-selecting objects more likely to be affected by secular
dynamics.

Asteroid families interacting with secular resonances can also be 
identified in domains of proper frequencies rather than proper elements.
In the Hierarchical Clustering Method of \citet{Zappala_1990},
asteroid families are identified with the following procedure: given an
individual asteroid the distance between this object and the other one is 
computed. If the distance is less than a threshold limit 
($d_{\rm cutoff}$), the new object is added to the list.
The procedure is repeated until no new family member is found.
A critical point in this procedure is related to the choice of
a reasonable metric in the three-dimensional element space. In
\citet{Zappala_1990}, the distance is defined as

\begin{equation}
 d_1=na\sqrt{k_1({\Delta a}/a)^2+k_2(\Delta e)^2+k_3(\Delta \sin i)^2},
\label{eq: stan_metr}
\end{equation}

\noindent where $n$ is the asteroid mean motion; $\Delta x$ the difference in
proper $a$, $e$, and $\sin i$; and $k_1, k_2, k_3$ are weighting factors,
defined as $k_1 = 5/4, k_2 = 2, k_3 = 2$. Other choices
of weighting factors are possible and yield similar results.
\citet{Carruba_2007} looked for families that interacted with $g+s$-type
secular resonances in a $(n,g,g+s)$, such as the Eos domain using a distance
metric of the form:

\begin{equation}
 d_2=\sqrt{h_1({\Delta n}/h_0)^2+h_2(\Delta g)^2+h_3(\Delta (g+s))^2},
\label{eq: freq_metr}
\end{equation}

\noindent where $h_0$ is a normalization factor of dimension 1 degree/arcsec,
and the simplest choice for the $h_i$ (i = 1, 3) weights is to take them
all equal to 1. The distance in frequency space then has the units
of arcsec/yr. The method looking for families in frequency domains was called
Frequency Hierarchical Clustering Method, of FHCM, by these authors.
Families found in this domain were, for appropriate choices
of $d_2$, able to connect to objects
that drifted in secular resonances of $g+s$-type and that were not recognized
as family members by the traditional HCM in proper element domain.

\begin{figure*}
  \centering
  \begin{minipage}[c]{0.49\textwidth}
    \centering \includegraphics[width=3.1in]{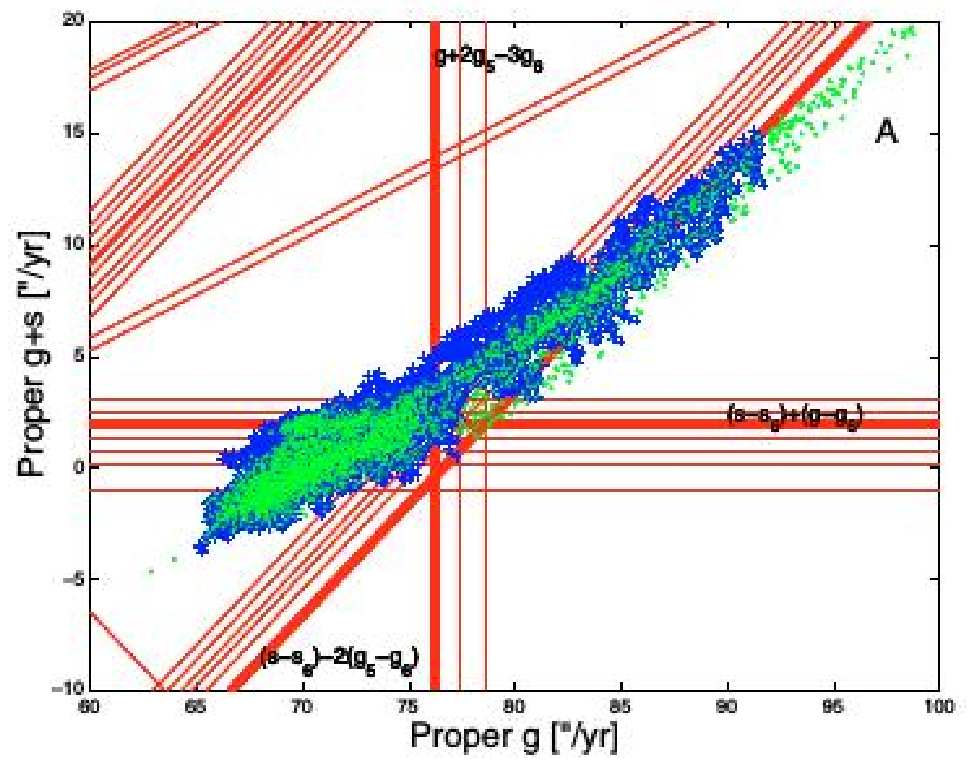}
  \end{minipage}%
  \begin{minipage}[c]{0.49\textwidth}
    \centering \includegraphics[width=3.1in]{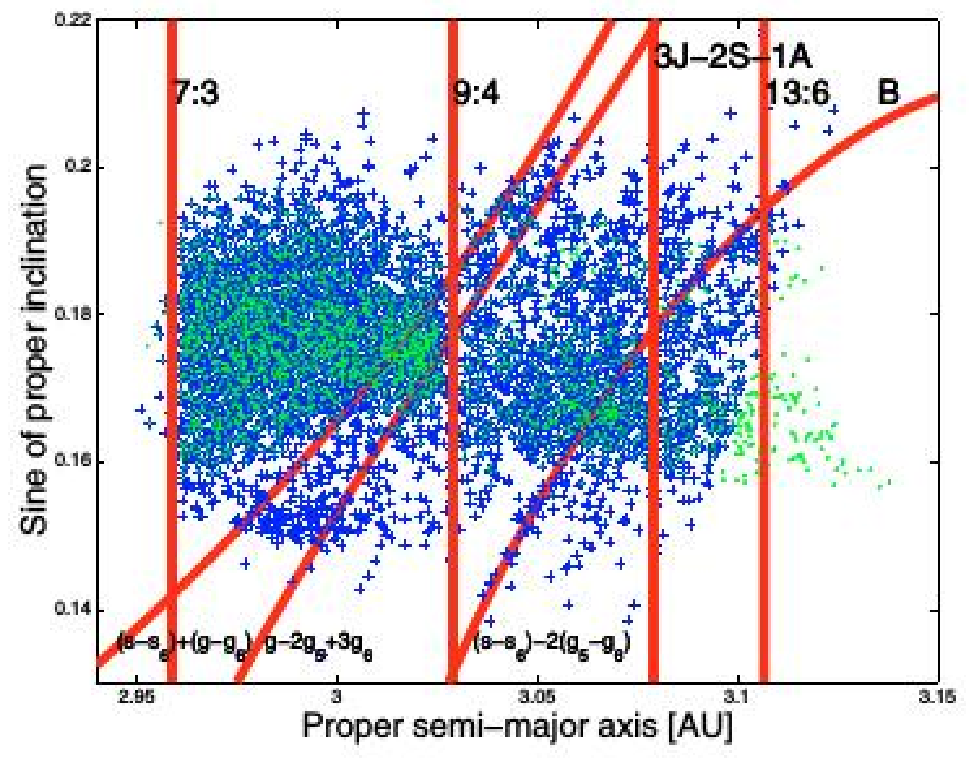}
  \end{minipage}
\caption{A $(g,g+s)$ (panel A) and a $(a,\sin i)$ (panel B)
projection of the Eos family obtained with the classical HCM (green dots),
and with the frequency HCM (blue crosses), as from Figure 6 of 
\citet{Carruba_2007}.}
\label{fig: eos}
\end{figure*}

Fig.~\ref{fig: eos} displays a $(g,g+s)$ (panel A) and a $(a,\sin i)$
(panel B) projection of the Eos family obtained with the classical HCM
(green dots), and with the frequency HCM (blue crosses), as from
\citet{Carruba_2007}. Secular resonances up to order 6, i.e., those
resonances for which the sum of the coefficients in the resonant
argument does not exceed 6, are shown in the
$(a,\sin i)$ plane and are identified by thicker lines in the $(g, g+s)$ plane.
Among other results, the Eos family identified in the HCM domain was
able to connect to four asteroids (20845) 2000 UY102, (21211) 1994
PP36, (33780) 1999 RU171, and (62948) 2000 VE32, currently
inside the $z_1$ resonance and at low values of proper $e$ and $\sin i$.
\citet{Vokrouhlicky_2006a} hypothesized that those objects were
former members of the Eos family that diffused to their current
position due to the interplay of the Yarkovsky effect and the $z_1$
resonance, and this was later confirmed by subsequent taxonomic analysis
of these asteroids that showed that these objects are of the same, peculiar
type of most of the Eos family asteroids, the K type, and therefore
most likely to have originated from this group. Other
objects diffusing in $g$, $s$ and other types of secular resonances were
also identified by the FHCM, but not by the standard HCM.

Other families interacting with other types of secular resonances,
such as $s$, $g$, $g+s$, $g-s$, $2g+s$, and $3g+s$, and the appropriate
distance metrics to study each case were also investigated in
\citet{Carruba_2007, Carruba_2009a}.  Interested readers can found additional
information in those papers.

\section{Families interacting with first-order secular resonances}
\label{sec: first-order res}

Once asteroids interacting with secular resonances have been identified,
using methods described in the previous section, information from
secular dynamics can then be used to set constraints on the dynamical
evolution of asteroid families interacting with secular resonances.  Here
we will review some of the results of the last decade on families in linear
and non-linear secular resonances.

\begin{figure}
\centering
\centering \includegraphics [width=0.45\textwidth]{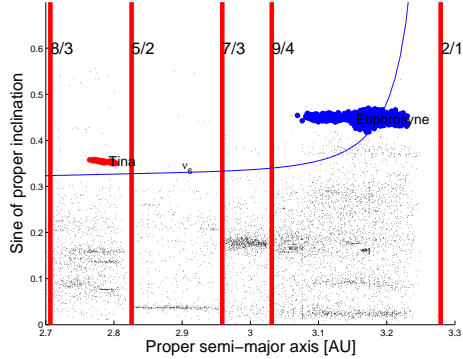}

\caption{The orbital location in the $(a,\sin i)$ plane of the Tina
  (red dots) and Euphrosyne (blue dots) families.  The location of the
  ${\nu}_6$ resonance was computed for the eccentricity value of (1222) Tina
  by using the second-order and fourth-degree secular perturbation theory of
  \citet{Milani_1992}. Vertical red lines displays the location of the
  main mean-motion resonances in the region.}
\label{Fig: nu6_fam}
\end{figure}

Among linear resonances, the ${\nu}_6 = g-g_6$ resonance is one of the main
effective mechanisms for increasing an asteroid eccentricity, and one of
the main source of NEA \citep{Morbidelli_1991, Bottke_2002, Granvik_2016}.
This resonance interacts with the Tina \citep{Carruba_2011} and
Euphrosyne \citep{Carruba_2014} families, and sets the boundary for
highly inclined objects in the central and outer main belt
\citep{Carruba_2010}.  Fig.~\ref{Fig: nu6_fam} displays an $(a,\sin i)$
projection of the location of the Tina and Euphrosyne families.   

\begin{figure}
\centering
\centering \includegraphics [width=0.45\textwidth]{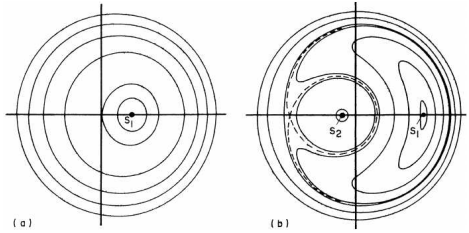}

\caption{The phase space in the $(e\cos{\sigma},e\sin{\sigma})$ domain
  of a first-order resonance far (panel a) and near (panel b) the perturbing
objects. Figure adapted from \citet{Ferraz_1985}.}
\label{Fig: ferraz}
\end{figure}

The resonance topology depends on the distance of the asteroid with
respect to Saturn.  Fig.~\ref{Fig: ferraz} displays the phase space
of equi-Hamiltonian curves for first-order resonances, as shown in the
classical work of \citet{Ferraz_1985}.  For first order resonances, at
higher distances from the perturber, which for the ${\nu}_6$ resonance would
correspond to Saturn and the inner main belt, the phase space of the resonance
would have an equilibrium point at $0^{\circ}$, and two possible
types of orbits: libration around the equilibrium point, for
which the resonant argument of the resonance will oscillates near
$0^{\circ}$, and circulation, for which the resonant argument
will cover all possible range of values.  The curve separating these
two classes of orbits is called the separatrix.  This type of dynamical
behavior is shown in Fig.~\ref{Fig: ferraz}, panel A.
Closer to the perturbing planet, however, another point of equilibrium
at $180^{\circ}$ may appear, and the separatrix may form a loop around
this equilibrium point.  Apart from orbit of libration near $0^{\circ}$
and circulation, a new class of orbits may be possible in these
regions:  the resonant argument may librate around $180^{\circ}$.
\citet{Ferraz_1985} calls this type of orbits paradoxical librators,
while other authors \citep{Morbidelli_1991} define them as anti-aligned
librators. The latter name is justified by the fact that, for the
${\nu}_6$ resonance, the argument of pericenter of Saturn and of the
perturbed object are roughly separated by $180^{\circ}$.
These orbits are not really on libration, but rather are
circulating orbits trapped by the separatrix loop.

\begin{figure*}
  \centering
  \begin{minipage}[c]{0.49\textwidth}
    \centering \includegraphics[width=3.1in]{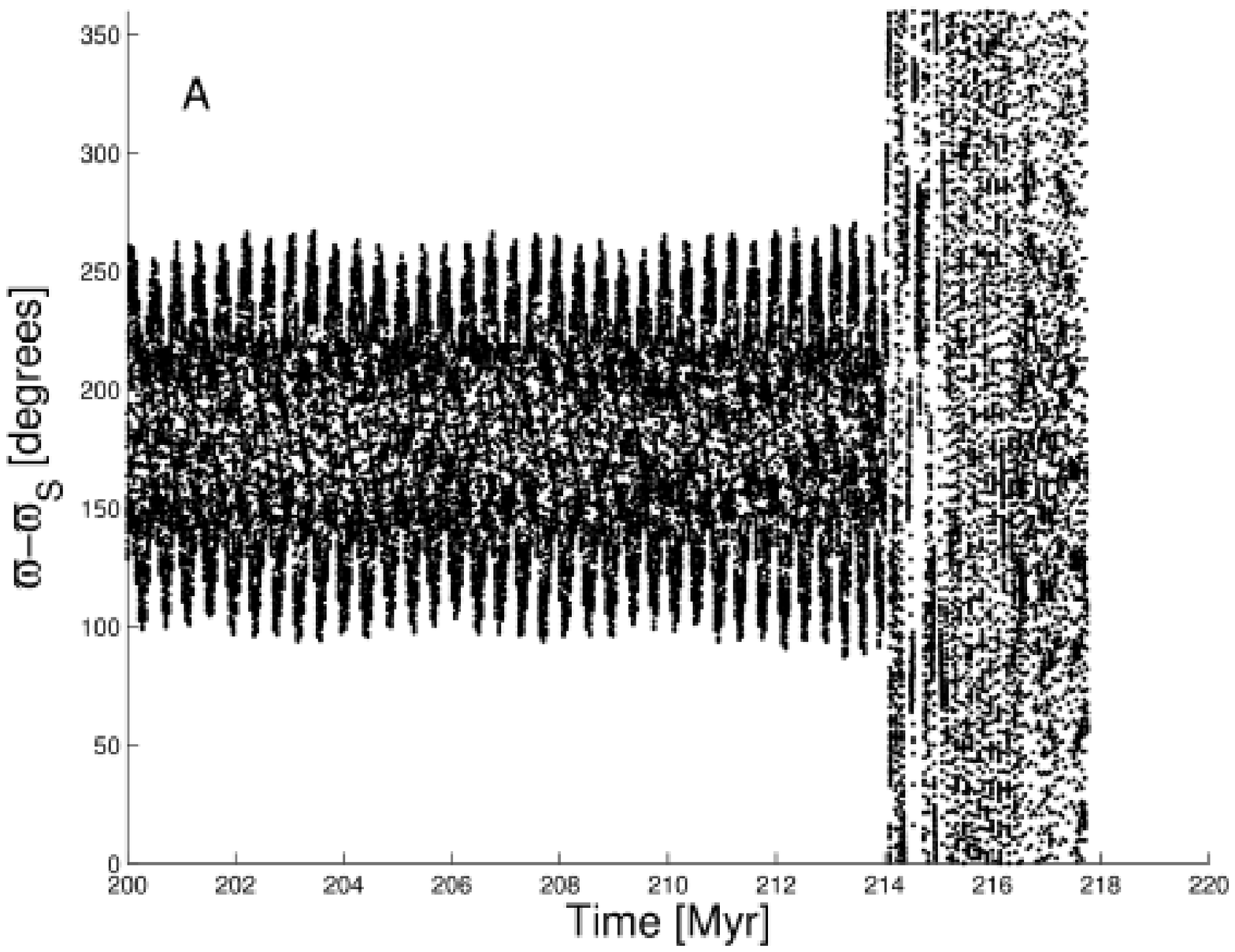}
  \end{minipage}%
  \begin{minipage}[c]{0.49\textwidth}
    \centering \includegraphics[width=3.1in]{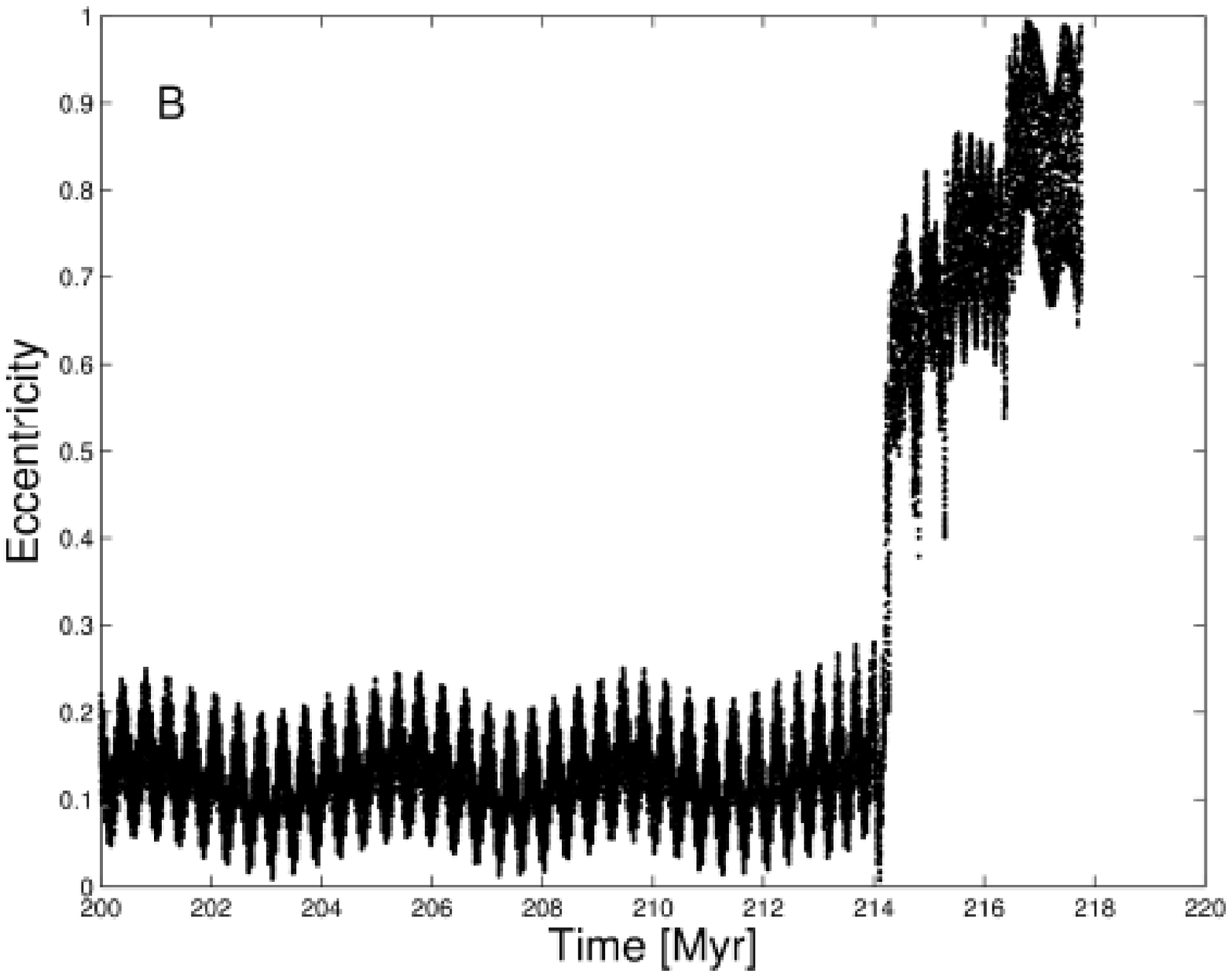}
  \end{minipage}
  \caption{Panel A: the time-evolution of the osculating resonance argument
    for a clone of the asteroid (211412) (2002 VL103). Panel B: the
    time-evolution of the osculating eccentricity of the same test particle.
    From Fig.~7 of \citet{Carruba_2011}.}
\label{Fig: Tina_ecc}
\end{figure*}

The Tina family is the only asteroid family currently known
whose members are all in anti-aligned states of the ${\nu}_6$ secular
resonance \citep{Carruba_2011}.
This orbital configuration protects the family members from reaching
planetary-crossing orbits.  Since anti-aligned orbits cannot cross
the separatrix, the maximum eccentricity that an orbit can reach is limited,
and may not reach planet-crossing levels.  This behavior is shown in
Fig.~\ref{Fig: Tina_ecc} for a clone of the asteroid (211412) (2002 VL103),
a member of the Tina family.  As long as the asteroid remains inside
the anti-aligned librating configuration, oscillations in eccentricity are
limited.  Once the asteroid escapes the stable region and goes into a
circulating orbit, changes in eccentricity grows dramatically and
$e$ reaches Mars-crossing levels of 0.4 in short time-scales.
All Tina family members are therefore found in a stable island of the
${\nu}_6$ secular resonance.

\begin{figure}
\centering
\centering \includegraphics [width=0.45\textwidth]{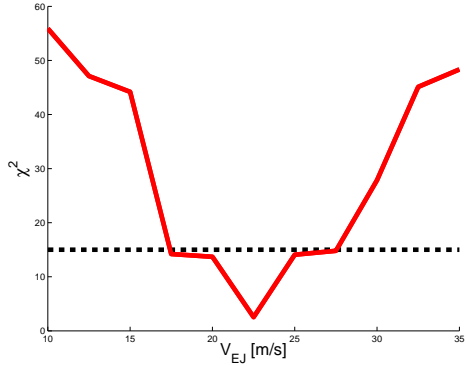}

\caption{The values of ${\chi}_2$ of the synthetic families as a function
  of the values of $V_{\rm EJ}$.  The horizontal line displays the probability
  confidence level. From Fig.~11 of \citet{Carruba_2011}.}
\label{Fig: chi2VEJ}
\end{figure}

The resonant nature of the Tina family allows to obtain information on the
original ejection velocity field of this group, not available for
families not affected by secular resonances.  At the simplest level of
perturbation theory, the ${\nu}_6$ resonance is characterized by the
conservation of the quantities $K_1 = \sqrt{a}$ and $K_2=\sqrt{a(1-e^2)}
(1-\cos i)$.  The quantity:

\begin{equation}
  K_2^{'}=\frac{K_2}{K_1}=\sqrt{1-e^2}\,(1-\cos i),
\label{eq: k2}
\end{equation}

\noindent
is preserved even when the Yarkovsky force is accounted for
\citep{Carruba_2011}. The current distribution of the $K_2^{'}$ quantity,
shown in Fig.~\ref{Fig: K2} for Tina asteroids, therefore preserves
information on the original one.  If we assume that
the ejection velocity field that created a Tina family was isotropic,
Gaussian, centered around 0 and with a size-dependent standard deviation
that follows the relationship:

\begin{figure}
\centering
\centering \includegraphics [width=0.45\textwidth]{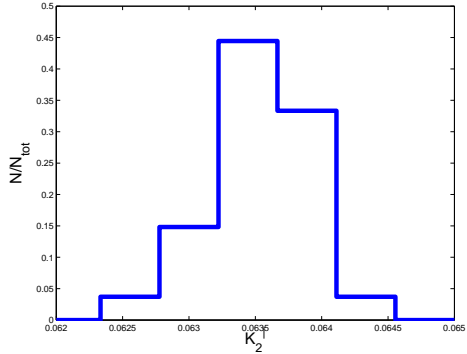}

\caption{The current distribution of the $K_2^{'}$ quantity for Tina family
  members. From Fig.~10 of \citet{Carruba_2011}.}
\label{Fig: K2}
\end{figure}

\begin{equation}
 V_{\rm SD}=V_{\rm EJ} \left(\frac{5\, {\rm km}}{D}\right),
\label{eq: V_SD}
\end{equation}

\noindent where $D$ is the asteroid diameter in km, and $V_{\rm EJ}$ is
a free parameter, usually of the order of the estimated escape velocity from
the parent body, we can then best-fit the
currently observed $K_2^{'}$ values with those of simulated asteroid
families with different $V_{\rm EJ}$ parameters.  For each family, we introduce
a ${\chi}^2$-like variable defined as:

\begin{equation}
  {\chi}^2=\sum_{i=1}^{N_{\rm int}} \frac{(q_i-p_i)^2}{q_i},
  \label{eq: chi2}
\end{equation}

\noindent where $N_{\rm int}$ is the number of interval used for the values of
$K_2^{'}$, $q_i$ is the number of real objects in the i-th interval in
$K_2^{'}$ and $p_i$ is the number of synthetic family members in the same i-th
interval. Fig.~\ref{Fig: chi2VEJ} shows how the value of ${\chi}_2$ changes
with the value of $V_{\rm EJ}$ adopted to generate the synthetic family.
The minimal value of ${\chi}_2$ (best fit) is obtained for
$V_{\rm EJ}=22\pm5$ m/s, in good agreement with the $20.0\pm2.5$ m/s estimate
obtained from Yarko-YORP methods \citep{Vokrouhlicky_2006c}. This method can
also be applied to families in non-linear secular resonances, which will be
the subject of the next section of this paper.

\section{Families in non-linear secular resonances}
\label{sec: non-linear res}

\begin{figure}
\centering
\centering \includegraphics [width=0.45\textwidth]{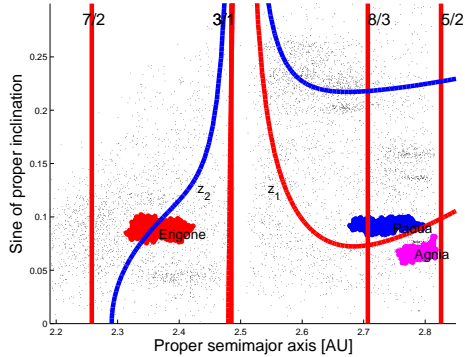}

\caption{The orbital location of the Padua, Agnia, and Erigone family and
  their position with respect to the $z_1$ and $z_2$ secular resonances.}
\label{Fig: fam_nonlin}
\end{figure}

There are asteroid families interacting with non-linear secular resonances of
the $z_k$ sequence. More than 75\% of the members of the Agnia
\citep{Vokrouhlicky_2006b} and Padua \citep{Carruba_2009} families are on
librating states of the $z_1$ =$g-g_6+s-s_6$ resonance, while 14\%
of the members of the Erigone family are in librating states of
the $z_2=2(g-g_6)+s-s_6$ resonance \citep{Carruba_2015}.  The Agnia family,
which is now know to have the sub-family of (3395) Jitka \citep{Spoto_2015},
was the first group identified to have the majority of its members in
librating states of a non-linear secular resonance.
The orbital position of these families and their relative location with respect
to the appropriate $z_k$ resonances is shown in Fig.~\ref{Fig: fam_nonlin}.

It is possible to show that the $K_2^{'}(z_1)$ quantity below is preserved for
the $z_1$ secular resonance \citep{Vokrouhlicky_2006b, Carruba_2009}:

\begin{equation}
  K_2^{'}=\frac{K_2}{K_1}=\sqrt{1-e^2}\,(2-\cos i),
\label{eq: k2_z1}
\end{equation}

By performing an analysis similar to that discussed in
Sect.~\ref{sec: first-order res} for the Tina family it can be shown that
the value of the ejection velocity parameter compatible with the $K_2^{'}(z_1)$
distribution is $15\pm 5$ m/s for the Agnia
family \citep{Vokrouhlicky_2006b} and $35\pm 8$ m/s for the Padua
one \citep{Carruba_2009} .

Concerning other families, \citet{Carruba_2015} showed that the 
$K_2^{'}(z_2) = K_2^{'}=K_2/K_1=\sqrt{1-e^2}\,(3-\cos i)$
quantity is not preserved for asteroids members of the Erigone family.
Yet, it is possible to obtain information on the age of this group from
secular dynamics.  Assuming that the current population of $z_2$
librators is in steady-state, one can use this information to set lower
limits on the Erigone family age.

\begin{figure}
  \centering
  \centering \includegraphics [width=0.45\textwidth]{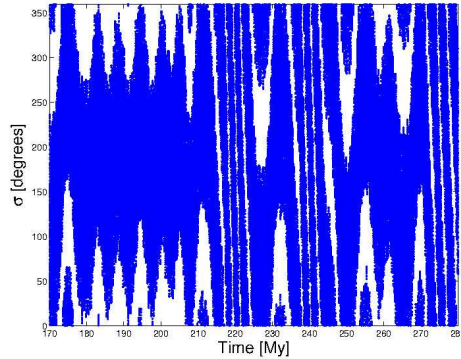}

\caption{Time behavior of the $z_2$ resonant argument of a simulated
particle between 170 and 280 My. From Fig.~11 of \citet{Carruba_2015}.}
\label{fig: res_ang_long}
\end{figure}

\citet{Carruba_2015} generated a fictitious Erigone family with optimal
$V_{\rm EJ}$ parameter, as found with Yarko-YORP methods, and integrated this
group over 400 My,  beyond the maximum possible value of the age of this
group. They then analyzed the resonant argument $\sigma$ of the $z_2$ secular
resonance for all simulated particles.  The behavior of this angle could 
be quite complex: because of the drift in semi-major axis caused by 
non-gravitational forces, asteroids can be alternate between times in which they
are captured into resonance, and times in which they escape to circulating
orbits, and vice-versa.  Fig.~\ref{fig: res_ang_long} displays the time
dependence of the $z_2$ resonant argument of a simulated particle that
alternated between phases of libration and circulation in the $z_2$ secular
resonance.  Overall, while the number of $z_2$ librators 
changes with time, if the population of librators 
from the Erigone family is in a steady-state, this number
should fluctuate around the median value, with fluctuation of the
order of one standard deviation.  A similar behavior was 
observed for the population of asteroids currently inside the
M2:1A mean-motion resonance \citep{Gallardo_2011}.  The minimum
time needed to reach a steady-state could therefore be used
to set lower limits on the age of the Erigone family

\begin{figure}
\centering
\centering \includegraphics [width=0.45\textwidth]{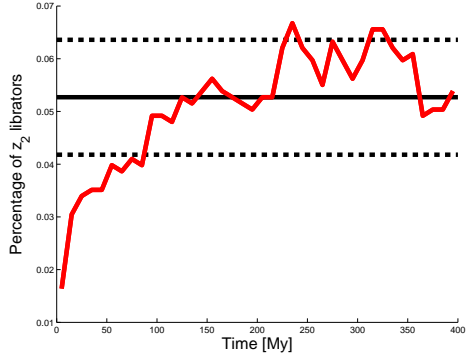}

\caption{Fraction of simulated Erigone family members in $z_2$ librating 
states as a function of time, normalized with respect to the median 
value.  The horizontal blue line displays the 
median percentage of objects in the $z_2$ states, dashed lines display
levels of median fraction plus or minus its standard deviation.
From Fig.~12 of \citet{Carruba_2015}.}
\label{fig: z2_dyn_constr}
\end{figure}

\citet{Carruba_2015} analyzed the resonant angle of all simulated
particles, and computed the fraction of family members in $z_2$ resonant
states as a function of time.  Fig.~\ref{fig: z2_dyn_constr} displays the
results: the number of $z_2$ librators fluctuates with time, but reaches its
median value after 125 My (after $\simeq$ 90 My if we consider
the median value plus or minus the standard deviations as an estimate
of the error, blue dashed lines in Fig.~\ref{fig: z2_dyn_constr}).  
This sets a lower limit on the family age, compatible with estimates
from the literature that yield an age older than 150 Myr for this group.

Recently, a new class of secular resonances, involving Ceres and
massive asteroids as pertubers, has been identified \citep{Novakovic_2015}.
Their effect on the dynamical evolution of asteroid families will be
discussed in the next section.

\section{Secular resonances with Ceres and massive asteroids}
\label{sec: ceres-res}

\begin{figure}
  \centering
  \centering \includegraphics [width=0.45\textwidth]{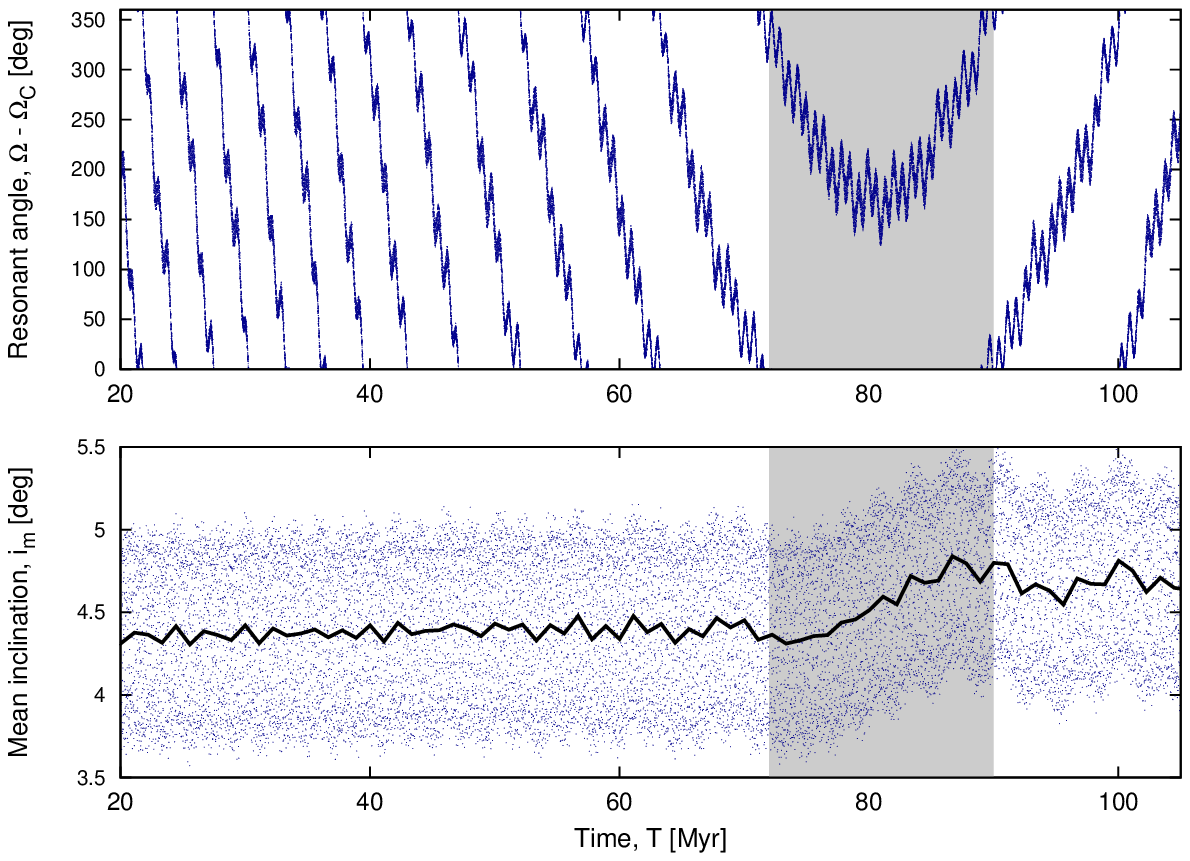}

\caption{Time behavior of the $\nu_{1C}$ resonant argument (top panel) and of
  the mean orbital inclination (bottom panel), of a simulated particle located
  in the region of the Hoffmeister family. The time period in which the mean
  inclination is increasing for about $0.5$~degrees is clearly related to the
  interval when the resonant argument is librating (shaded area). The solid
  black line denotes the average of the mean inclination to better appreciate
  the evolution.}
\label{Fig: nu1c}
\end{figure}

The importance of secular resonances with massive asteroids has been recently
realized by \citet{Novakovic_2015}. In particular, these authors found that
the spreading in orbital inclination seen in the Hoffmeister family, is a
consequence of the nodal liner secular resonance with Ceres, namely
$\nu_{1C}$ = $s-s_{C}$. It was actually shown that passing through the
$\nu_{1C}$ resonance may cause significant changes in the orbital inclination
of an asteroid.  This effect is visible in Fig.~\ref{Fig: nu1c},
where the time evolution of the inclination and of the resonant angle of the
$\nu_{1C}$ resonance, for an asteroid belonging to the Hoffmeister family, are
plotted.  The instant of increase in inclination corresponds very well to
the libration period of the resonant angle.

This was the first direct proof that a secular resonance 
between Ceres and other asteroids, prior completely overlooked, can cause 
significant orbital evolution of the latter. The result opened several
new possibilities to investigate the dynamics of small bodies, and initiates
a series of related studies.

\begin{figure*}
  \centering
  \begin{minipage}[c]{0.49\textwidth}
    \centering \includegraphics[width=3.1in]{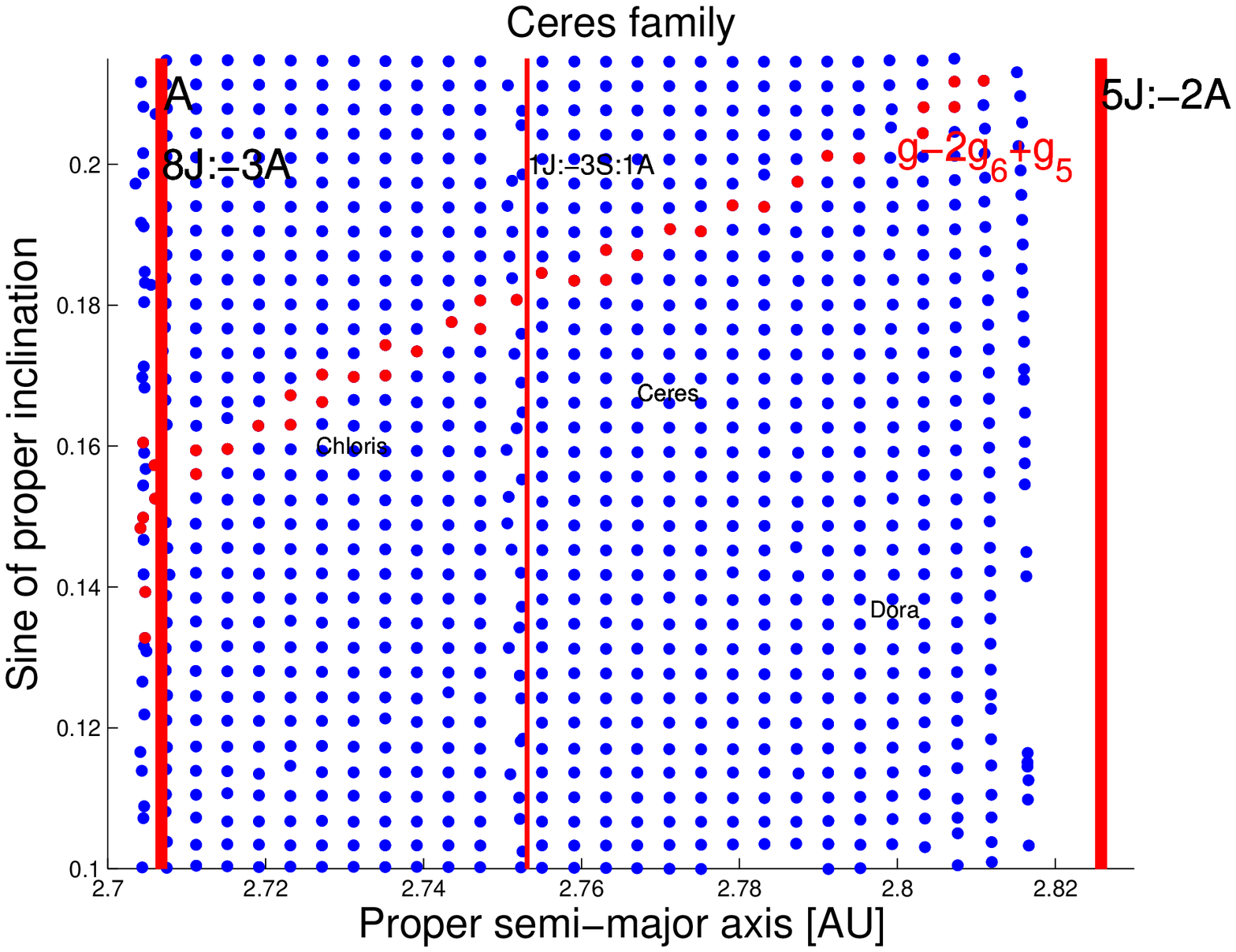}
  \end{minipage}%
  \begin{minipage}[c]{0.49\textwidth}
    \centering \includegraphics[width=3.1in]{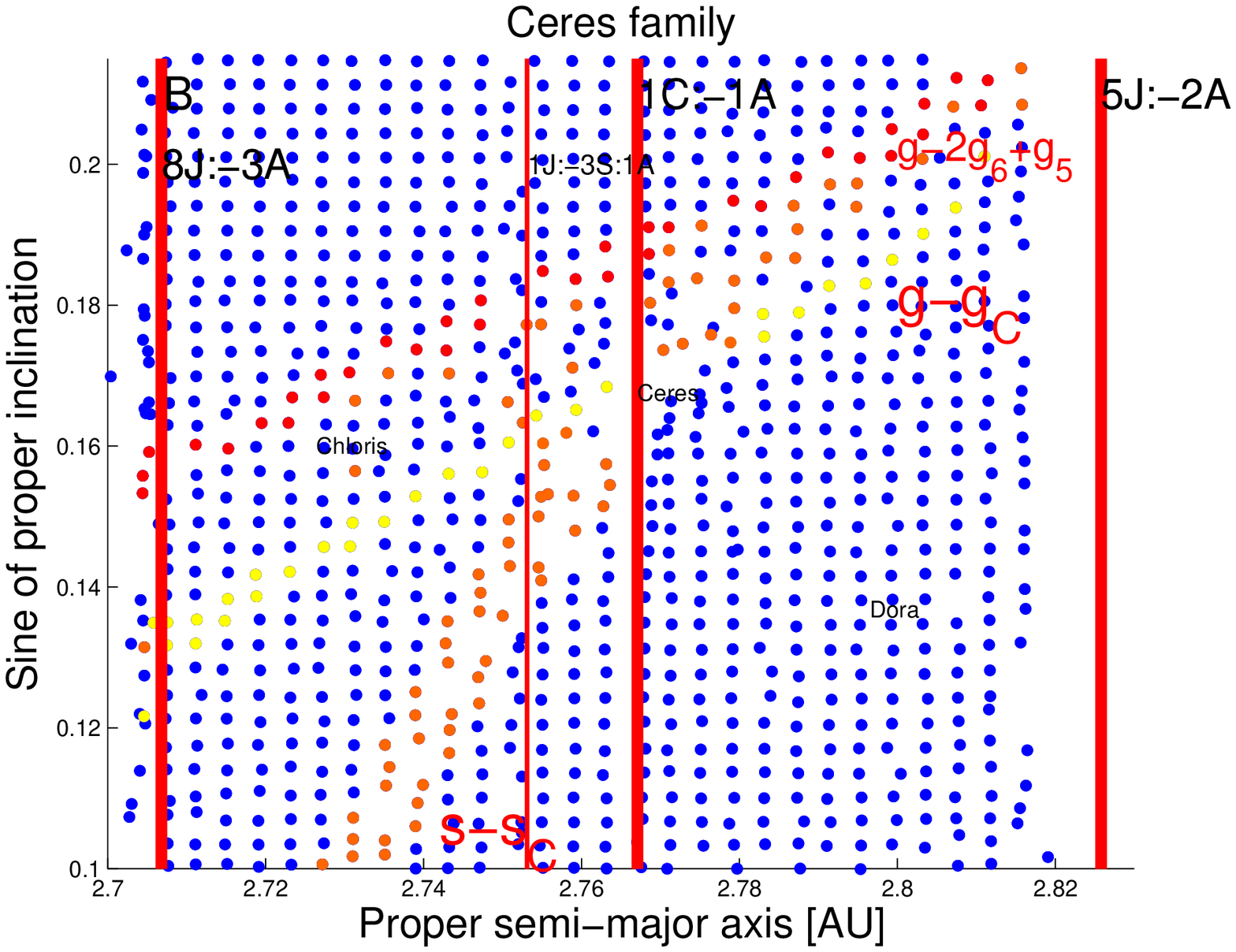}
  \end{minipage}
  \caption{Dynamical maps for the orbital region of Ceres obtained by
    integrating test particles under the influence of all planets
    (panel A), and all planets and Ceres as a massive body (panel B).
    Unstable regions associated with mean-motion resonances appear as
    vertical strips. Secular resonance appear as inclined bands of
    aligned dots. Dynamically stable regions are shown as uniformly
    covered by blue dots. Vertical lines display the location of the main
    mean-motion resonances in the area. Red filled dots in both panels
    show the locations of 'likely resonators' in the $g - 2g_6 + g_5$
    secular resonance. Yellow and orange filled dots in panel B show the
    orbital locations of likely resonators in the linear secular resonances
    with Ceres of pericenter and node, respectively. The orbital location of
    1 Ceres, 668 Dora, and 410 Chloris are labeled.  Adapted from Fig.~1
    of \citet{Carruba_2016}.}
\label{Fig: ceres_map}
\end{figure*}

One of the most interesting works in this respect was performed by
\citet{Carruba_2016}. The authors showed that secular resonances with Ceres
may explain why there is no Ceres family, showing that methods for identifying
asteroid family based on a search of neighbors pairs, like the HCM, cannot
reveal the possible existence of a Ceres family. In order to highlight a
role of Ceres, \citet{Carruba_2016} obtained synthetic proper elements
dynamical map for the region of the central main belt without and with Ceres
as a perturber.  Results are shown in Fig.~\ref{Fig: ceres_map}.

As observed in Fig.~\ref{Fig: ceres_map}, panel B, accounting for
Ceres causes the appearance of a 1:1 mean-motion resonance with this
dwarf planet. Most importantly, linear secular resonances of nodes
$s - s_C$ and pericenter $g - g_C$, first detected by \citet{Novakovic_2015},
significantly destabilize the orbits in the proximity of Ceres. Combined with
the long-term effect of close encounters with Ceres, this has interesting
consequences for the survival of members of the Ceres family
in the central main belt. Not many family members are expected
to survive near Ceres, and this would cause significant difficulties
in using standard dynamical family identification techniques, since
they are based on looking for pairs of neighbors in proper element
domains. Since the close neighbors of Ceres would
have been removed on a short time-scale, only objects whose distance from
Ceres is higher than the average distance between pairs
of asteroids in the central main belt would have survived.   Results
of simulations of fictitious Ceres families in \citet{Carruba_2016}
showed that secular dynamics would indeed clear the region near Ceres of
neighbors, making it impossible to identify a Ceres family after timescales
of 350 Myr.

Motivated by the aforementioned results, \citet{Tsirvoulis_2016} made a map
of the locations of the linear secular resonances with Ceres and Vesta in the
asteroid belt, and investigated the magnitude of the perturbations. Their
results show that in some cases the strength of secular resonances with Ceres
and Vesta is similar to that of non-linear secular resonances with the major
planets. \citet{Tsirvoulis_2016} have also identified several asteroid
families crossed by the secular resonances 
with Ceres and Vesta, and pointed out that the post-impact evolution of these
families may be significantly affected by aforementioned resonances.
The list of these families includes the Astrid, Hoffmeister, Seinajoki and some 
other groups \citep[see also][]{Novakovic_2016}.

Secular dynamics with massive bodies can provide constraints on the ages of
the ``$v_W$ leptokurtic families'', or families characterized by their
interaction with secular resonances with Ceres, which will be the subject of
the next section.

\section{$v_W$ leptokurtic asteroid families}
\label{sec: vw}

The change in inclination of a family members is related to the
perpendicular component of the velocity at infinity $v_W$ through the
corresponding Gauss equation:

\begin{equation}
\delta i =\frac{\sqrt{1-e^2}}{na}\frac{\cos{(\omega+f)}}{1+e\cos{f}}\,\delta v_W,
\label{eq: gauss_i}
\end{equation}

\noindent 
Where the changes in proper $\delta i$ are computed with respect to the family
center of mass, and $f$ and $\omega$ are the (generally unknown) true
anomaly and perihelion argument of the disrupted body at the time of impact.
If a family interacts with a secular resonance of node with a massive
body, its inclination distribution will become more peaked and with
larger tails, when compared with a Gaussian one.
As a consequence, the value of the Pearson kurtosis of the $v_W$ component
of the ejection velocity field, ${\gamma}_2(v_W)$, will also increase.
\citep{Carruba_2016b} listed all families characterized by large
values of this parameter. The minimum time needed to attain the current
value of ${\gamma}_2(v_W)$ for the $v_W$ leptokurtic families can be used to
set constraints on their ages.  This method for dating $v_W$ leptokurtic
families has been used, so far, to date six asteroid families, those of
Astrid, Hoffmeister, Gallia, Barcelona, Hansa, and Rafita.

\begin{figure*}
  \centering
  \begin{minipage}[c]{0.49\textwidth}
    \centering \includegraphics[width=3.1in]{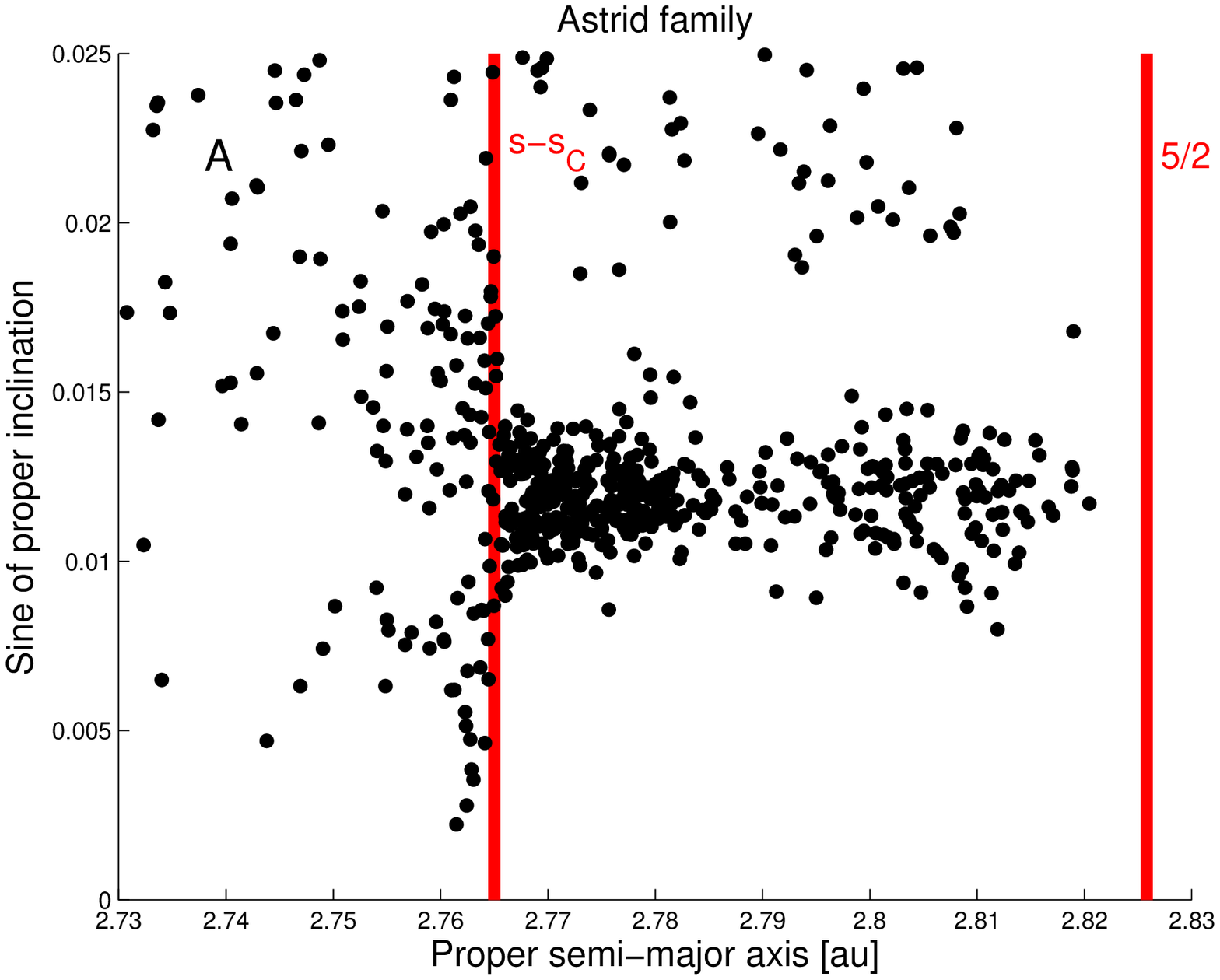}
  \end{minipage}%
  \begin{minipage}[c]{0.49\textwidth}
    \centering \includegraphics[width=3.1in]{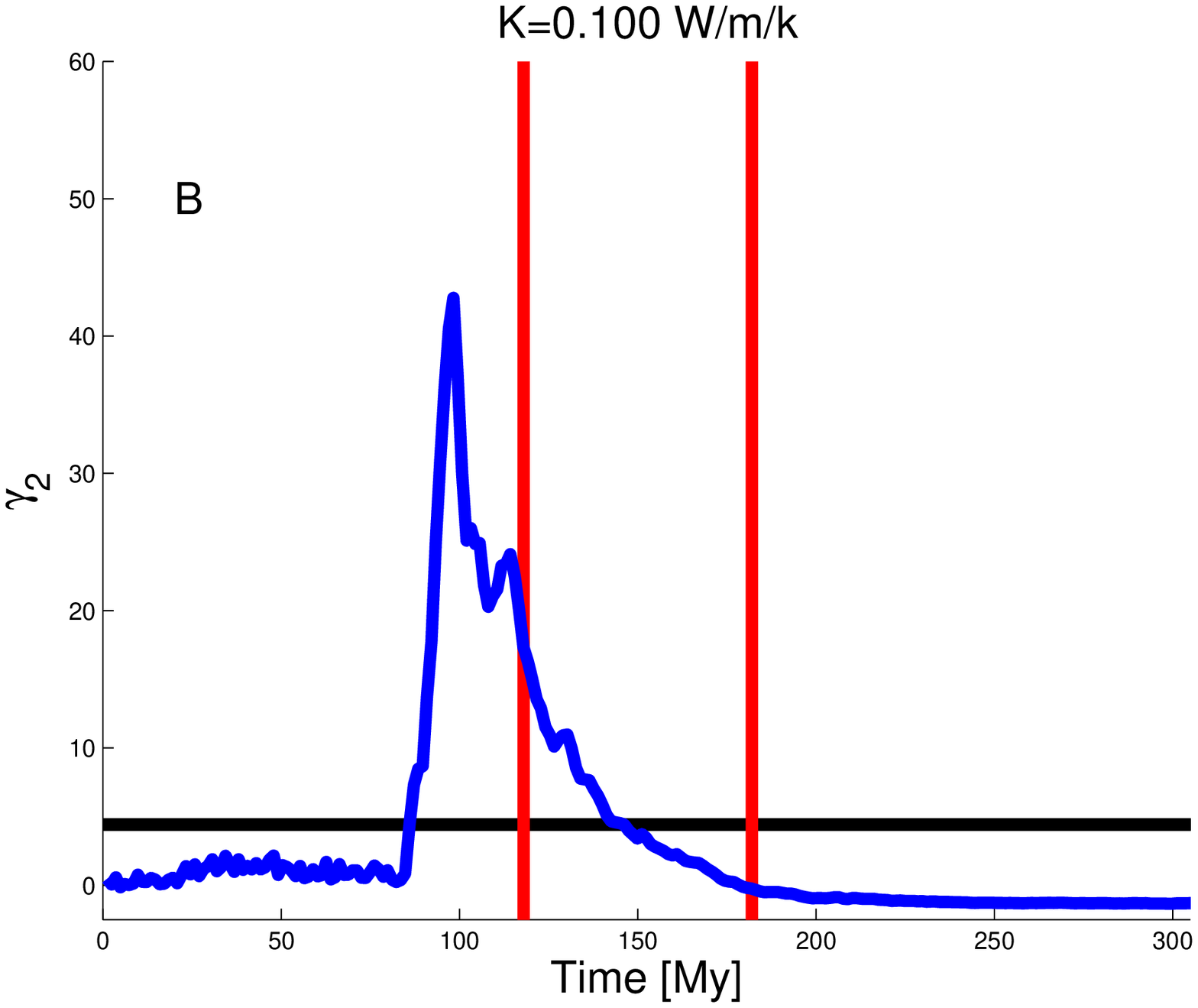}
  \end{minipage}
  \caption{Panel A: an $(a,\sin{i})$ projection of members of the Astrid
    family.  Vertical red lines display the location of local secular and
    mean-motion resonances. Panel B: the time evolution of the
    ${\gamma}_2(v_W)$ for a family with the optimal parameters of the Yarkovsky
    force. Vertical red lines displays the estimated age of the Astrid family,
    while the horizontal black line shows the current value of
    ${\gamma}_2(v_W)$ for this family. Adapted from figures of
    \citet{Carruba_2016c}.}
\label{Fig: Astrid}
\end{figure*}

The Astrid family was the first family for which this method was applied
\citep{Carruba_2016c}. This family, characterized by a squid-like
appearance in the $(a,\sin i)$ domain (see Fig.~\ref{Fig: Astrid},
panel A), interacts with the $s-s_C$ resonance, which is responsible
for its distribution in inclination.

By simulating various fictitious families and by demanding that the
current value of the ${\gamma}_2(v_W)$ of the distribution in be reached over
the estimated lifetime of the family of $140\pm 30$ Myr, \citet{Carruba_2016c}
showed that the thermal conductivity of Astrid family members should be
$0.001$ W m$^{-1}$ K$^{-1}$, unusual for a C-type family, and that the
surface and bulk density should be higher than 1000 kg m$^{-3}$. 
The time evolution of the ${\gamma}_2(v_W)$ for a family with the optimal
parameters of the Yarkovsky force can be seen in Fig.~\ref{Fig: Astrid},
panel B.

\begin{figure*}
  \centering
  \begin{minipage}[c]{0.49\textwidth}
    \centering \includegraphics[width=3.1in]{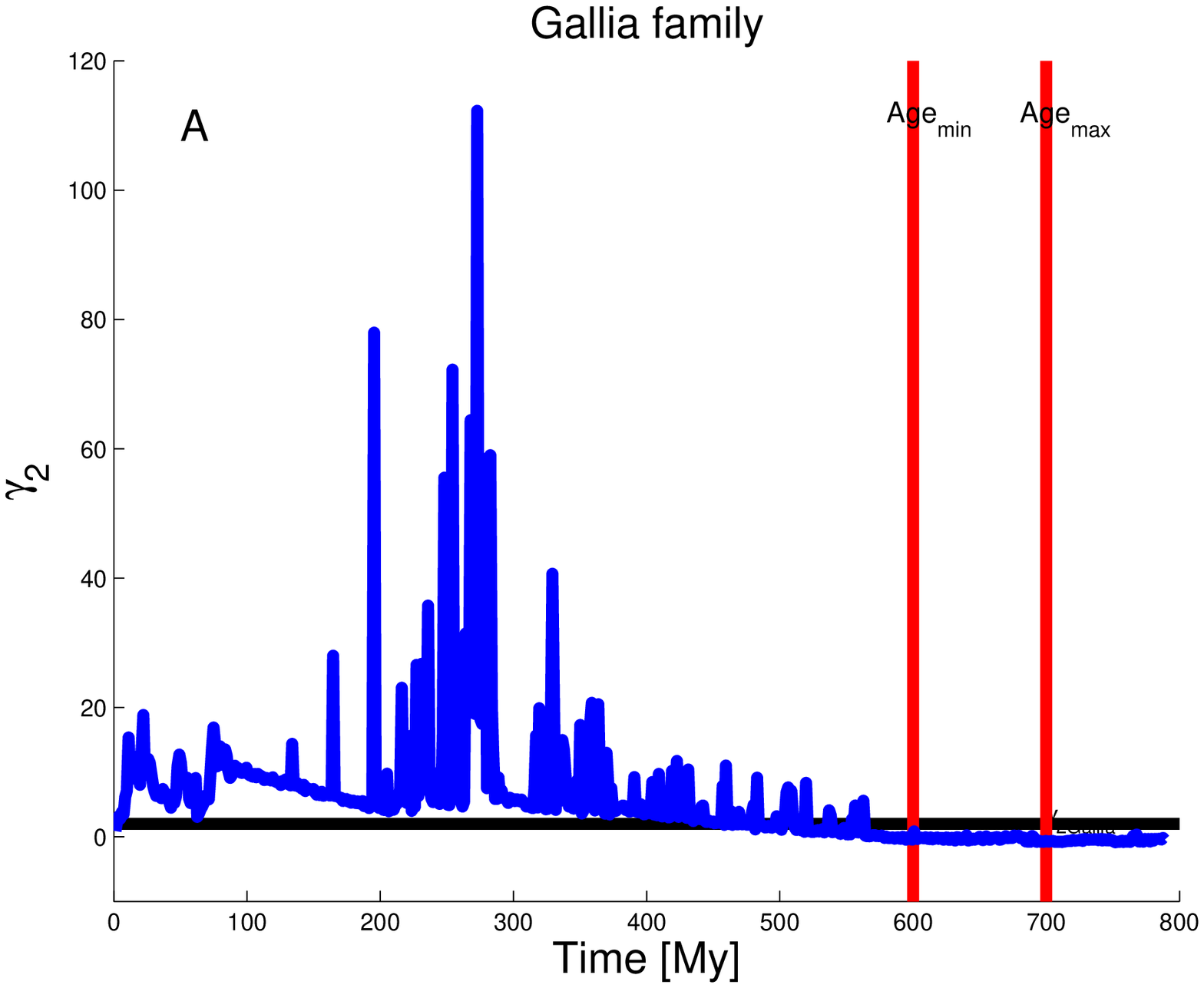}
  \end{minipage}%
  \begin{minipage}[c]{0.49\textwidth}
    \centering \includegraphics[width=3.1in]{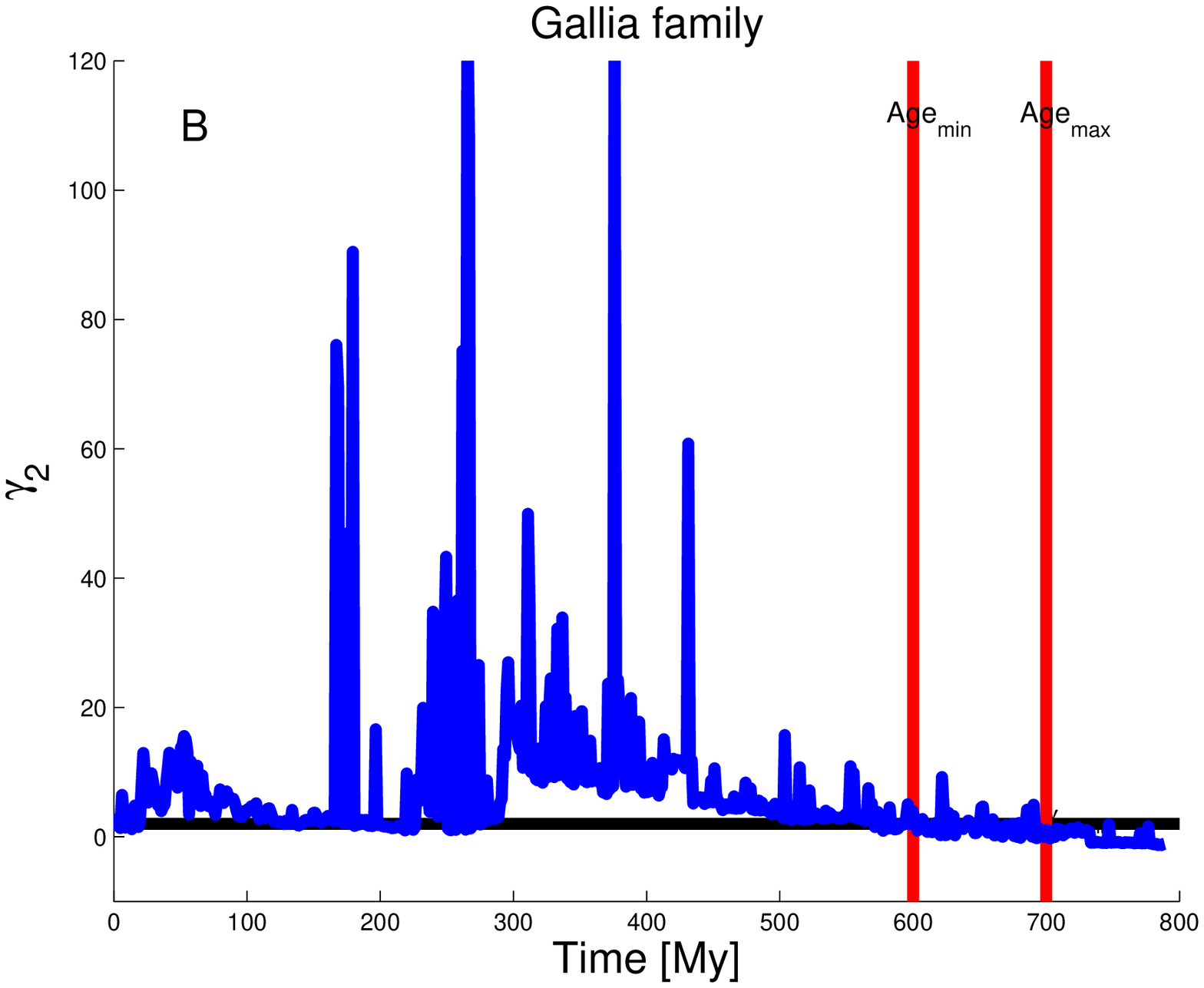}
  \end{minipage}
  \caption{Panel A: the time evolution of the ${\gamma}_2(v_W)$ parameter
    for a simulation without Ceres as a massive perturber. Panel B displays
    the same, but for a simulation that also account for Ceres as a
    massive body. Vertical red lines displays the estimated age of the
    Gallia family. The horizontal line show the current value of
    ${\gamma}_2(v_W)$.  Adapted from figures of \citet{Carruba_2016d}.}
\label{Fig: Gallia}
\end{figure*}

Three of the most $v_W$ leptokurtic families are found in the highly inclined
region of the central main belt: the Hansa, the Gallia, and the Barcelona
families.  \citet{Carruba_2016d} used the ${\gamma}_2(v_W)$ approach
to obtain family ages for these three families, whose results are summarized
in Table~\ref{Table: vW}. For the Gallia family, the current value of
${\gamma}_2(v_W)$ can only be reached over the estimated family age if Ceres is
considered as a massive perturber.  Fig.~\ref{Fig: Gallia} shows the time
behavior of this parameter for a scenario without (panel A) and with (panel B)
Ceres as a massive body.  Indeed, current values of ${\gamma}_2(v_W)$ can be
reached over the estimated family age only if Ceres is considered as a massive
body.

Finally, although this was not the main focus of their work,
\citet{Aljbaae_2016} applied the $v_W$ approach of family dating to the
Rafita asteroid family, increasing the total number of families with age
estimated in this way to five. All the results obtained so far with this
family dating method are summarized in Table~\ref{Table: vW}.

\begin{table}
\begin{center}
  \caption{Age estimates and values of the $V_{EJ}$ parameter for the families
  for which the $v_W$ approach has been used so far.  The first column
  displays the Family Identification Number (FIN), as from
  \citet{Nesvorny_2015}, the second column the family name, the third
  the current value of ${\gamma}_2(v_W)$, as from \citet{Carruba_2016b},
  the fourth the estimated family age, and the fifth the estimated value
  of the $V_{EJ}$ parameter.}
\label{Table: vW}
\begin{tabular}{|c|c|c|c|c|}
\hline
      &        &                   &            &             \\ 
FIN   & Family &  Current          & Estimated  & Estimated    \\
      & Name   & ${\gamma}_2(v_W)$ & age [Myr] & $V_{\rm EJ}$ [m/s] \\
      &        &                   &           &               \\
\hline 
      &        &                   &           &               \\
515 & 1128 Astrid   & 3.69 & $140\pm30$        & $10^{+30}_{-10}$ \\
518 & 1644 Rafita   & 0.63 & $480\pm15$        & $20^{+30}_{-15}$ \\
802 & 148 Gallia    & 2.02 & $630^{+30}_{-70}$   & $5^{+17}_{-5}$  \\
803 & 480 Hansa     & 0.81 & $460^{+280}_{-360}$ & $80^{+10}_{-65}$ \\
805 & 945 Barcelona & 1.48 & $265^{+45}_{-35}$   & $15^{+20}_{-15}$ \\
      &        &           &                   &               \\
\hline
\end{tabular}
\end{center}
\end{table}

\section{Conclusions}
\label{sec: conc}

In this review paper, we saw how:

\begin{itemize}

\item Secular resonances have a complex structure in the proper
  $(a,e,\sin i)$ domain.   But they appear as lines in proper
  $(n,g,s)$ domains.  In particular $g$-type resonances should be plotted
  in a domain in which the y-axis is the $g$ frequency, $s$-type
  resonances will appear as horizontal lines in a plane in which the
  $s$ frequency is on the y-axis, and non-linear secular resonances
  should be plotted in a domain appropriate to their type ($g+s$ for
  $(g+s)$-type resonances, $g-s$ for $(g-s)$ resonances, etc.). 
  By selecting a cutoff value near the combination of planetary frequencies
  for each resonance, it is possible to select asteroids more likely
  to be affected by secular dynamics, the so-called "likely resonators".
  Asteroid families interacting with secular resonances can also be
  identified in domains of proper frequencies
  \citep{Carruba_2007,Carruba_2009a}.

\item Because of their tilted shape in the space of proper orbital elements,
  secular resonances may effectively change proper eccentricity and/or
  inclination if the proper semi-major axis is evolving due to the Yarkovsky
  effect. This unique property has led to setting evidence of Yarkovsky
  effect influence in a number of asteroid families.
  
\item Because of the preservation of quantities associated with the local
  secular dynamics, asteroid families interacting with linear (Tina family and
  the ${\nu}_6$ secular resonance) and non-linear (Agnia and Padua family and
  the $z_1$ resonance, the Erigone and the $z_2$ resonance) secular resonances
  still preserve information on their original ejection velocity field.
  The study of this families provides clues on the mechanisms of family
  formation not available for other, non-resonant families.

\item The interaction of asteroid families with nodal secular resonances
  with Ceres may significantly change their inclination distribution, making
  it more leptokurtic, i.e., more peaked and with larger tails, when
  compared with a Gaussian one.  By modeling the time evolution of the
  $v_W$ component of the ejection velocity field, it is possible to set
  constraint on the family age and original ejection velocity field of
  families for which the current value of the Pearson Kurtosis of $v_W$
  is significantly larger than 0.  So far, six asteroids families, those
  of Astrid, Hoffmeister, Gallia, Barcelona, Hansa, and Rafita, have been
  dated with this method.
\end{itemize}

Overall, secular dynamics can provide invaluable hints for our understanding of
the dynamical evolution of asteroid families, in many cases not available
for non-resonant groups.  This could be of potential great interest for
the many new smaller families recently identified in \citet{Nesvorny_2015}
and \citet{Milani_2014}, whose resonant nature has yet to be investigated.

\section*{Acknowledgments}
We are grateful to the two reviewers of this paper, an anonymous
reviewer and Dr. Federica Spoto, for comments and suggestions that
significantly improved the quality of this work.
We would like to thank the S\~{a}o Paulo State Science Foundation 
(FAPESP) that supported this work via the grant 16/04476-8,
and the Brazilian National Research Council (CNPq, grant 312313/2014-4).
DV's work was funded by the Czech Science Foundation through the grant
GA13-01308S. BN acknowledges support by the Ministry of Education, Science and
Technological Development of the Republic of Serbia, project 176011.
We acknowledge the use of data from the Asteroid Dynamics Site
(AstDys) \\ 
(http://hamilton.dm.unipi.it/astdys, \citet{Knezevic_2003}).

%This publication makes use of data products from the Wide-field 
%Infrared Survey Explorer (WISE) and Near-Earth Objects (NEOWISE), which
%are a joint project of the University of California, Los Angeles, and the
%Jet Propulsion Laboratory/California Institute of Technology, funded by the
%National Aeronautics and Space Administration.  


\begin{thebibliography}{}

\bibitem[Aljbaae et~al. (2016)]{Aljbaae_2016} Aljbaae, S., Carruba, V.,
 Masiero, J., Domingos, R.C., Huaman, M., 2016. The Rafita asteroid family.
 MNRAS 467, 1016--1023.

\bibitem[Bottke et al. (2001)]{Bottke_2001} Bottke, W.F., Vokrouhlick\'{y},
 D., Bro\v{z}, M., Nesvorn\'{y}, D., Morbidelli, A., 2001. Dynamical
 spreading of asteroid families via the Yarkovsky effect: The Koronis family
 and beyond. Science 294, 1693--1696.

\bibitem[Bottke et al. (2002)]{Bottke_2002} Bottke, W.F., Morbidelli, A.,
  Jedicke, R., Petit, J.-M., Levison, H.F., Michel, P., Metcalfe, T.S.,
  2002. Debiased orbital and absolute magnitude distribution of the
  near-Earth objects. Icarus 156, 399--433.
 
\bibitem[Brouwer and van Woerkom (1950)]{Brouwer_1950} Brouwer, D.,
  van Woerkom, A.J.J., 1950. Astronomical papers prepared for the use of
  the American ephemeris and nautical almanac, Vol.~13, Washington:
  U.S.~Govt.~Print.~Off., pp.~81--107.

\bibitem[Brouwer (1951)]{Brouwer_1951} Brouwer, D., 1951. Secular variations
 of the orbital elements of minor planets. AJ 56, 9--32.

\bibitem[Carruba et al. (2005)]{Carruba_2005} Carruba, V., Michtchenko, T. A.,
 Roig, F., Ferraz-Mello, S., Nesvorn\'{y}, D., 2005. On the V-type asteroids
 outside the Vesta family. I. Interplay of nonlinear secular resonances and
 the Yarkovsky effect: the cases of 956 Elisa and 809 Lundia.
 A\&A 441, 819--829.
  
\bibitem[Carruba and Michtchenko (2007)]{Carruba_2007} Carruba, V., Michtchenko
 T., 2007. frequency approach to identifying asteroid families. A\&A 475,
 1145--1158.
  
\bibitem[Carruba and Michtchenko (2009)]{Carruba_2009a} Carruba, V., Michtchenko
 T., 2009. A frequency approach to identifying asteroid families. II. Families
 interacting with nonlinear secular resonances and low-order mean-motion
 resonances. A\&A 493, 267--282.
  
\bibitem[Carruba (2009)]{Carruba_2009} Carruba, V. 2009. The (not so) peculiar
 case of the Padua family. MNRAS 395, 358--377.

\bibitem[Carruba (2010)]{Carruba_2010} Carruba, V. 2010. The stable archipelago
 in the region of the Pallas and Hansa dynamical families. MNRAS 408, 580--600.

\bibitem[Carruba and Morbidelli (2011)]{Carruba_2011} Carruba, V., Morbidelli,
 A., 2011. On the first $\nu_6$ anti-aligned librating asteroid family of
 Tina. MNRAS 412, 2040--2051.
  
\bibitem[Carruba et al. (2014a)]{Carruba_2014} Carruba, V., Aljbaae, S.,
  Souami, D., 2014a. Peculiar Euphrosyne. ApJ 792, 46.
  
\bibitem[Carruba et al. (2014b)]{Carruba_2014b} Carruba, V., Huaman, M.,
 Domingos, R.C., Dos Santos, C.R., Souami, D., 2014b. Dynamical evolution of
 V-type asteroids in the central main belt. MNRAS 439, 3168--3179.
  
\bibitem[Carruba et al. (2015)]{Carruba_2015} Carruba, V., Aljbaae, S., Winter,
 O.C., 2015. On the Erigone family and the $z_2$ secular resonance. MNRAS
 455, 2279--2288.

\bibitem[Carruba et al. (2016a)]{Carruba_2016} Carruba, V., Nesvorn\'{y}, D.,
 Marchi, S., Aljbaae, S., 2016a. Footprints of a possible Ceres asteroid
 paleo-family. MNRAS 458, 1117--1126.

\bibitem[Carruba and Nesvorn\'{y} (2016)]{Carruba_2016b} Carruba, V.,
  Nesvorn\'{y}, D., 2016. Constraints on the original ejection velocity
  fields of asteroid families. MNRAS 457, 1332--1338.

\bibitem[Carruba (2016)]{Carruba_2016c} Carruba, V., 2016. On the Astrid
 asteroid family. MNRAS 461, 1605--1613.

\bibitem[Carruba et al. (2016b)]{Carruba_2016d} Carruba, V., Nesvorn\'{y}, D.,
 Domingos, R.C., Aljbaae, S., Huaman, M., 2016b. On the highly inclined $v_W$
 leptokurtic asteroid families. MNRAS 463, 705--711.

\bibitem[Ferraz-Mello (1985)]{Ferraz_1985} Ferraz-Mello, S., 1985. Resonance
 in regular variables. I - Morphogenetic analysis of the orbits in the case
 of a first-order resonance. CMDA 35, 209--220. 
  
\bibitem[Ferraz-Mello (2007)]{Ferraz_2007} Ferraz-Mello, S., 2007. Canonical
 Perturbation Theories, Degenerate Systems and Resonances. Springer, New York.

\bibitem[Gallardo et al. (2011)]{Gallardo_2011} Gallardo, T., Venturini, J.,
 Roig, F., Gil-Hutton, R., 2011. Origin and sustainability of the population of
 asteroids captured in the exterior resonance 1:2 with Mars. Icarus 214,
 632--644.

\bibitem[Granvik et al. (2016)]{Granvik_2016} Granvik, M., Morbidelli, A.,
 Jedicke, R., Bolin, B., Bottke, W.F., Beshore, E., Vokrouhlick\'y, D.,
 Delb\`o, M., Michel, P., 2016. Super-catastrophic disruption of asteroids
 at small perihelion distances. Nature 530, 303--306.

\bibitem[Hirayama (1923)]{Hirayama_1923} Hirayama, K., 1923. Annales
  de l'Observatoire Astronomique de Tokyo, 11.
  
\bibitem[Kne\v{z}evi\'{c} et al. (1991)]{Knezevic_1991} Kne\v{z}evi\'{c}, Z.,
 Milani, A., Farinella, P., Froeschle, C., Froeschle, Ch., 1991. Secular
 resonances from 2 to 50 AU. Icarus 93, 316--330.

\bibitem[Kne\v{z}evi\'{c} and Milani (2000)]{Knezevic_2000} Kne\v{z}evi\'{c},
  Z., Milani, A., 2000. Synthetic Proper Elements for Outer Main Belt
  Asteroids.  CMDA 78, pp. 17--46.

\bibitem[Kne\v{z}evi\'{c} and Milani (2003)]{Knezevic_2003} Kne\v{z}evi\'{c},
 Z., Milani, A., 2003. Proper element catalogs and asteroid families. A\&A
 403, 1165--1173.

\bibitem[Lemaitre (1994)]{Lemaitre_1994b} Lemaitre, A., 1994.
 Hungaria: A potential new family. In: Seventy-five years of Hirayama
 asteroid families: The role of collisions in the solar system history
 (Y. Kozai, R.P. Binzel \& T. Hirayama eds.) Astronomical Society of the
 Pacific Conference Series, Vol. 63, pp. 140--145.

\bibitem[Lemaitre and Morbidelli (1994)]{Lemaitre_1994a} Lemaitre, A.,
 Morbidelli, A., 1994. Proper elements for highly inclined asteroidal orbits.
 CMDA 60, 29--56.
 
\bibitem[Machuca and Carruba (2012)]{Machuca_2012} Machuca, J. F., Carruba, V.,
  2012. Secular dynamics and family identification among highly inclined
  asteroids in the Euphrosyne region. MNRAS 420, 1779--1798.
  
\bibitem[Milani and Knezevic (1990)]{Milani_1990} Milani, A., Kne\v{z}evi\'c,
 Z., 1990. Secular perturbation theory and computation of asteroid proper
 elements. CMDA 49, 347--411. 

\bibitem[Milani and Kne\v{z}evi\'{c} (1992)]{Milani_1992} Milani, A.,
 Kne\v{z}evi\'{c}, Z., 1992. Asteroid proper elements and secular resonances.
 Icarus 98, 211--232.

\bibitem[Milani and Kne\v{z}evi\'{c} (1994)]{Milani_1994} Milani, A.,
  Kne\v{z}evi\'{c}, Z., 1994. Asteroid proper elements and the dynamical
  structure of the asteroid main belt. Icarus 107, 219--254.
  
\bibitem[Milani et al. (2010)]{Milani_2010} Milani, A., Kne\v{z}evi\'{c}, Z.,
 Novakovi\'{c}, B., Cellino, A., 2010. Dynamics of the Hungaria asteroids.
 Icarus 207, 769--794.
  
\bibitem[Milani et al. (2014)]{Milani_2014} Milani, A., Cellino, A.,
 Kne{\v z}evi{\'c}, Z., Novakovi{\'c}, B., Spoto, F., Paolicchi, P.,
 2014. Asteroid families classification: Exploiting very large datasets.
 Icarus 239, 46--73.

\bibitem[Milani et al. (2017)]{Milani_2017} Milani, A., Kne{\v z}evi{\'c}, Z.,
 Spoto, F., Cellino, A., Novakovi\'c, B., Tsirvoulis, G., 2017. On the ages
 of resonant, eroded and fossil asteroid families. Icarus 288, 240--264.

\bibitem[Mili{\'c} {\v Z}itnik and Novakovi{\'c} (2015)]{Zitnik_2015}
 Mili{\'c} {\v Z}itnik, I., Novakovi{\'c}, B., 2015. On some dynamical
 properties of the Phocaea region. MNRAS 451, 2109--2116.

\bibitem[Morbidelli and Henrard (1991)]{Morbidelli_1991} Morbidelli, A.,
  Henrard, J., 1991. The main secular resonances nu6, nu5 and nu16 in the
  asteroid belt. CMDA 51, 169--197.

\bibitem[Morbidelli (1993)]{Morbidelli_1993} Morbidelli, A., 1993. Asteroid
  secular resonant proper elements. Icarus 105, 48--66. 
  
\bibitem[Nesvorn\'{y} et al. (2015)]{Nesvorny_2015} Nesvorn\'{y}, D.,
 Bro\v{z}, M., Carruba, V., 2015. Identification and dynamical properties
 of asteroid families. in Asteroids~IV, (P. Michel, F. E. DeMeo,
 W. Bottke Eds.), University of Arizona Press, p.~297--321.

\bibitem[Novakovi{\'c} et al. (2015)]{Novakovic_2015} Novakovi{\'c}, B., 
 Maurel, C., Tsirvoulis, G., Kne{\v z}evi{\'c}, Z., 2015. Asteroid secular
 dynamics: Ceres' fingerprint identified. ApJL 807, L5.
 
\bibitem[Novakovi{\'c} et al. (2016)]{Novakovic_2016} Novakovi{\'c}, B.,
  Tsirvoulis, G., Mar{\`o}, S., Djo{\v s}ovi{\'c}, V., Maurel, C., 2016.
  Secular evolution of asteroid families: the role of Ceres. In the
  proceedings of the IAU Symposium 318 - Asteroids: New Observations,
  New Models (S. Chesley, A. Morbidelli, R. Jedicke \& D. Farnocchia eds.) 
  pp.~46--54. 
  
\bibitem[Spoto et al. (2015)]{Spoto_2015} Spoto, F., Milani, A.,
  Kne\v{z}evi\'{c}, Z., 2015. Asteroid family ages. Icarus 257, 275--289.

\bibitem[Tsirvoulis and Novakovi\'{c} (2016)]{Tsirvoulis_2016} Tsirvoulis, G., 
 Novakovi{\'c}, B., 2016. Secular resonances with Ceres and Vesta. Icarus 280,
 300--307.

\bibitem[Vokrouhlick\'{y} and Bro\v{z}(2002)]{Vokrouhlicky_2002}
  Vokrouhlick{\'y}, D., Bro\v{z}, M., (2002), Interaction of the
  Yarkovsky-drifting orbits with weak resonances: numerical evidence and
  challenges, in `Modern Celestial Mechanics: from Theory to Applications',
  eds. A. Celletti, S. Ferraz-Mello and J. Henrard (Kluwer Academic Publ,
  Dordrecht), pp. 467-472.
  
\bibitem[Vokrouhlick\'{y} et al. (2006a)]{Vokrouhlicky_2006a} Vokrouhlick{\'y},
 D., Bro{\v z}, M., Morbidelli, A., Bottke, W.F., Nesvorn\'y, D., Lazzaro, D.,
 Rivkin, A.S., 2006a. Yarkovsky footprints in the Eos family. Icarus 182,
 92--117. 

\bibitem[Vokrouhlick\'{y} et al. (2006b)]{Vokrouhlicky_2006b} Vokrouhlick\'{y}
 D., Bro\v{z}, M., Bottke, W.F., Nesvorn\'{y}, D., Morbidelli, A., 2006b.
 The peculiar case of the Agnia asteroid family. Icarus 183, 349--361.

\bibitem[Vokrouhlick\'{y} et al. (2006c)]{Vokrouhlicky_2006c} Vokrouhlick\'{y}
 D., Bro\v{z}, M., Bottke, W.F., Nesvorn\'{y}, D., Morbidelli, A., 2006c.
 Yarkovsky/YORP chronology of asteroid families. Icarus 182, 118--142.
  
\bibitem[Vokrouhlick\'{y} et al. (2010)]{Vokrouhlicky_2010} Vokrouhlick\'{y},
 D., Nesvorn\'{y}, D., Bottke, W.F., Morbidelli, A., 2010. Collisionally born
 family about 87 Sylvia. AJ 139, 2148--2158.

\bibitem[Vokrouhlick\'{y} et al. (2015)]{Vokrouhlicky_2015} Vokrouhlick\'{y},
 D., Bottke, W.F., Chesley, S.R., Scheeres, D.J., Statler, T.S., 2015.
 The Yarkovsky and YORP effects. in Asteroids~IV, (P. Michel, F. E. DeMeo,
 W.F. Bottke Eds.), University of Arizona Press, pp.~509--531.

\bibitem[Warner et al. (2009)]{Warner_2009} Warner, B. D., Harris, A.W.,
 Vokrouhlick\'{y}, D., Nesvorn\'{y}, D., Bottke, W.F., 2009. Analysis of the
 Hungaria population. Icarus 204, 172--182.

\bibitem[Williams (1969)]{Williams_1969} Williams, J.G., 1969. Secular
 Perturbations in the Solar System, Ph.D. Thesis, University of California,
 Los Angeles.

\bibitem[Yoshikawa (1987)]{Yoshikawa_1987} Yoshikawa, M., 1987. A simple
 analytical model for the secular resonance nu6 in the asteroidal belt.
 Celestial Mechanics 40, 233--272.

\bibitem[Zappal\`{a} et al. (1990)]{Zappala_1990} Zappal\`{a}, V.,
 Cellino, A., Farinella, P., Kne\v{z}evi\'{c}, Z., 1990. Asteroid families.
 I - Identification by hierarchical clustering and reliability assessment.
 AJ 100, 2030-2046.

\end{thebibliography}
\end{document}